# Assessing Executive Function Using a Computer Game: Computational Modeling of Cognitive Processes

Stuart Hagler, Holly B. Jimison, and Misha Pavel

*Abstract*—Early and reliable detection of cognitive decline is one of the most important challenges of current healthcare. In this project we developed an approach whereby a frequently played computer game can be used to assess a variety of cognitive processes and estimate the results of the pen-and paper Trail-Making-Test (TMT) – known to measure executive function, as well as visual pattern recognition, speed of processing, working memory, and set-switching ability. We developed a computational model of the TMT based on a decomposition of the test into several independent processes, each characterized by a set of parameters that can be estimated from play of a computer game designed to resemble the TMT. An empirical evaluation of the model suggests that it is possible to use the game data to estimate the parameters of the underlying cognitive processes and using the values of the parameters to estimate the TMT performance. Cognitive measures and trends in these measures can be used to identify individuals for further assessment, to provide a mechanism for improving the early detection of neurological problems, and to provide feedback and monitoring for cognitive interventions in the home.

*Index Terms*—Additive Stages, Computer Game, Executive Function, Fitts' Law, Neuropsychological Test

## I. Introduction

QUANTITATIVE assessment of cognitive function is an important component of caring for the aging as well as those with other dysfunctions such as traumatic brain injury and many other conditions affecting cognitive functions. The goal of this study is to find ways to assess and monitor subjects' cognitive performance in the subjects' home using information technology. In this paper, we show how a simple computer game in conjunction with computational model can be used for sensitive assessment and monitoring of components of executive function in individual subjects.

The computer game we consider in this paper bears a close relationship to a commonly administered, neuropsychological test – the (pen-and-paper) Trail Making Test (TMT). Typically administered as one test in a larger battery of tests, TMT is made up of two parts – TMT-A and TMT-B – each resembling a child's connect-the-dots puzzle. Each part, as with the puzzle, is completed by drawing a single, continuous line through all the "dots" in a specified order. The subject's score on each part of TMT is the time the subject took to draw the line to the last "dot." TMT is known to measure visuo-perceptual ability, working memory, and set-switching ability. [1, 2]

Computer-based implementations of neuropsychological tests, such as TMT, have potentially many advantages over traditional, pen-and-paper implementations, including: (1) uniformity of administration across subjects, and (2) more consistent scoring of performance. They also allow the possibility of decomposing performance on the test into performance on individual parts of the test. Researchers have examined the use of computer-based neuropsychological testing [3-5] and have found them to be promising for the cognitive assessment of older adults. [4, 5] In particular, computerized implementations of TMT have been developed (e.g. [6, 7]), however differences between performance on a computerized implementation of TMT and the standard pen-and-paper TMT, as measured by the scores on TMT-A and TMT-B, have been shown. [7] An alternative to simply implementing a computer-based TMT is to have the subject perform the pen-and-paper TMT while the test administrator notes the duration of the subject's moves the pen to each "dot" by selecting a button on a computer GUI each time a "dot" is selected. [8] This approach allows the performance on TMT to be decomposed into performance on each movement to a "dot."

Our approach is to focus on the time taken to make each move to each "dot" rather than on the time taken to draw the line through the whole set of "dots." Given the information about subject performance gained by examining all the individual moves to "dots" we can then estimate the time the subject would need to draw a line through a set of dots - such as those given on TMT. To obtain sufficiently accurate estimates of the underlying processes requires data for a large number of moves. To acquire the needed move data, we have constructed a simple computer game in which the subject

Manuscript received April 1, 2013; revised August 16, 2013. This work was supported by the National Institute on Standards and Technology's Advanced Technology Projects, by the Intel Corporation, the Alzheimer's Association and the National Institutes of Health (NIA grants P30AG024978 and ASMMI0116ST).

Stuart Hagler is with the Department of Biomedical Engineering, Oregon Health & Science University, Portland, OR 97239 USA (e-mail: haglers@ohsu.edu).

Holly B. Jimison is with College of Computer & Information Science / Bouvé College of Health Sciences, Northeastern University, Boston, MA 02115 USA (e-mail: h.jimison@neu.edu).

Misha Pavel is with College of Computer & Information Science / Bouvé College of Health Sciences, Northeastern University, Boston, MA 02115 USA (e-mail: m.pavel@neu.edu).

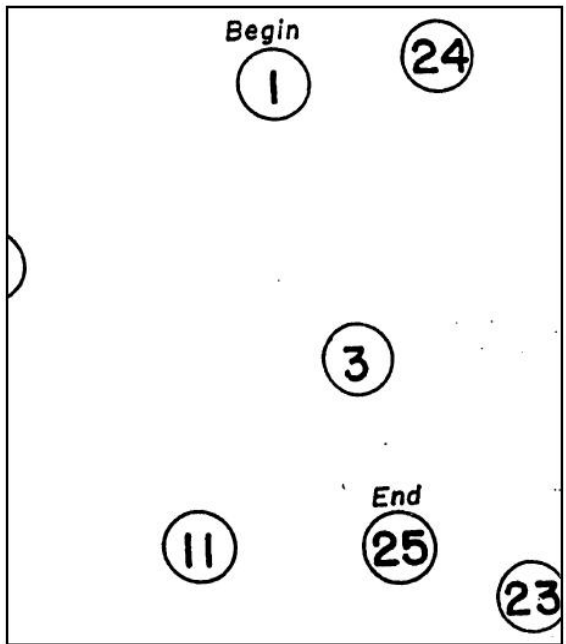

Fig. 1. A section of the TMT-A neuropsychological test. Note the words "Begin" and "End" indicating the locations of the first and last targets.

completes a series of rounds each of which consists of a set of randomly placed "dots" which the subject connects by using a computer mouse to select the "dots" in a specified sequence.

We develop a model for each move to a "dot" assuming a sequence of the three independent processes based on Donders' additive stages: [9, 10] recall, search, and motor. The motor stage, describing the movement of the pen or mouse from one "dot" to the next, is based on Fitts' law, characterizing rapid movements into specified target regions. [11-13]

## II. OVERVIEW OF TWO CONNECT-THE-DOTS TASKS

TMT is a pen-and-paper neuropsychological test that measures a subject's visuo-perceptual ability (ability to interpret visual information), working memory (ability to hold items in memory while completing a complex task), and set-switching ability (ability transition from a task involving one class of objects to a task involving a different class). [1, 2] Scavenger Hunt (SH) is a point-and-click, mouse-driven computer game with game mechanics designed to mimic the testing mechanics of TMT and yet be both challenging and fun so that people are willing to play it routinely over time in a home environment. Both TMT and SH are built around the *connect-the-dots task* that forms the basis of the child's puzzle giving the task its name. In this task, the subject must select a number of "dots" in sequence by drawing a line through them. We used both TMT and SH to validate our model of the connect-the-dots task given in Sec. III. The following sections provide an overview of TMT and SH.

We call the interval from the selection of one "dot" to the selection of the next "dot" a *move*.

### A. Trail Making Test

The pen-and-paper TMT is comprised of two separate tests: TMT-A and TMT-B. TMT-A and TMT-B are printed on a standard 8.5"x11" sheet of paper with 25 small (12mm diameter) circles, the *targets*, placed in a seemingly random pattern on the sheet. All targets in both tests have the same diameter or *width*, and contain a *label* which may a letter or a number. In TMT-A, a label is a number from 1 to 25, while in TMT-B a label is a letter from A to L or a number from 1 to 13. Labels only appear once on the test page. In addition, two targets on each test page are indicated by the presence of the words "Begin" and "End" near to (but outside of) these targets. TMT-A and TMT-B refer to test pages each with a specific arrangement of targets, and the same test pages are used every time TMT is administered. Fig. 1 shows a portion of TMT-A.

Prior to beginning TMT-A or TMT-B, the test administrator instructs the subject on how to correctly complete the test. This is done by walking the subject through a shorter - 8 target - sample test.

TMT-A and TMT-B each start with the subject being given the test page face down, the subject not having seen the test page. The test begins when the test administrator instructs the subject to start the test, and the subject turns over the test page and begins. The subject completes each test by drawing a single line, the *trail*, through all 25 targets in the specified order. In TMT-A, the targets are selected in ascending numerical order of the target labels (i.e. '1,2,3,…,24,25'), while in TMT-B, the order is ascending alphanumeric order of the target labels (i.e. '1,A,2,B,…,L,13'). The "Begin" is printed on the test page lies near the first target of the sequence and the "End" near the last.

The subject's performance on TMT is given in the form of a *score* on each of TMT-A and TMT-B, that is, the time taken to correctly draw a line through all targets on the test page in the specified order beginning when the test administrator instructs the subject to begin, and ending when the test administrator notes that the subject has reached the last target.

The test administrator also makes sure that the subject completes the test correctly, interrupting the test whenever the subject is observed to make an incorrect target selection (i.e. selecting a target out of sequence), as soon as the test administrator notices the error. Whenever such an *error* occurs, the test administrator instructs the subject to return to the last correctly selected target. Timing is not suspended during this process, and is included in the total time and thus in the test score, although, the number of errors for TMT-A and TMT-B are recorded by the test administrator. We call the sequence just outlined, the process of *recovering* from the error.

Guidelines for the administration of TMT are provided in [14]. TMT is normally administered in an office setting by a neuropsychologist once every 6 to 12 months to reduce practice effects associated with repeated completion of standardized tests. [15-17] Normative data for subjects with education in the range of the subject used in our study show TMT scores for ages 75-79 of 42 sec on TMT-A and 100 sec on TMT-B, and for ages 80-84 of 55 sec on TMT-A and 130 sec on TMT-B. [18] TMT is one of the most clinically useful neuropsychological tests and is routinely used in the diagnosis of many neurological conditions (Parkinson's, Alzheimer's and dementias, and general cognitive decline). [19] However, it is an expensive test and the infrequent assessment leads to delays in the detection of cognitive issues.

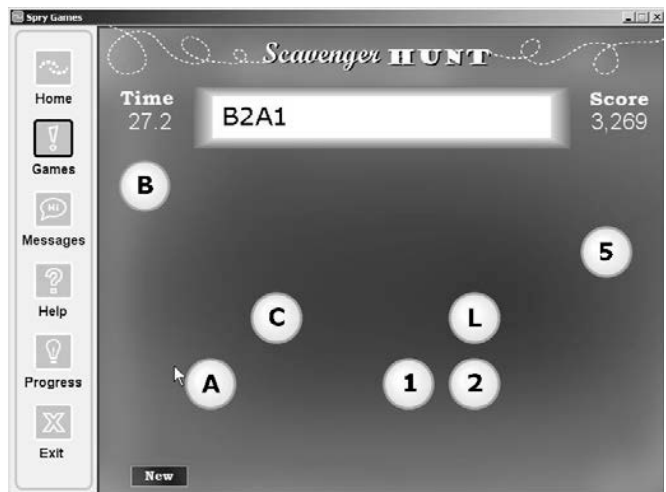

Fig. 2. A typical Scavenger Hunt round. The board for a round of Scavenger Hunt showing the time remaining in the game (27.2 sec), search string ('B2A1'), cumulative game score (3269), targets ('1','2','A','B'), and distractors ('5','C','L').

*B. Scavenger Hunt*

The point-and-click, mouse-driven SH computer game is intended to mimic the mechanics of TMT while presenting the subject with arbitrary target configurations rather than the two fixed target configurations present on TMT-A and TMT-B test pages. SH was designed to be more engaging and fun, and yet present cognitive tasks that would test cognitive functions similar to those of TMT. We have been able to demonstrate that older adults are able to learn the game and play it routinely on computers in their homes. [20]

The subject plays SH by completing a series of *rounds* each being a single connect-the-dots task. SH rounds must be completed in 30 sec (imposing a speed-accuracy tradeoff); if the subject fails to do so the round is lost and the game of SH ends. SH play continues from round to round until either a round is lost, or the subject elects to stop playing.

Fig. 2 shows a typical SH round. We call the large pane within the GUI containing the words "Scavenger Hunt" the *board*. The upper left hand corner of the board shows the amount of time left in the round. The upper right hand corner of the board shows the cumulative *score* for all the rounds completed so far. The box in the center of the upper center of the board contains the *search string* 'B,2,A,1'. The remainder of the board shows the array of *markers* for this round. The set of markers includes both *targets*: '1', '2', 'A', and 'B', and additional *distractors*: '5', 'C', and 'L'. The *number* of markers on the board for the round shown is 7. The subject would *play* this round of SH by using the computer mouse to connect-the-dots by *selecting* targets the targets 'B', '2', 'A', '1', in that order, by clicking on them. The *trail* made by the subject in SH is path taken by the mouse in the course of selecting the targets.

SH indicates correctly selected markers by coloring them green for the remainder of the round; no line indicating a trail is drawn. A subject makes an error when playing a round of SH by selecting any marker other than the one currently being looked for (i.e. the lowest unselected marker in the search string). SH indicates that an error has occurred by drawing a red "X" over the selected marker which remains until another marker is selected.

SH displays the subject's cumulative score on the game board for the rounds that have been completed. This score is used as feedback and motivation for the subject and not used to infer cognitive function or predict TMT scores. In this paper, we only refer to TMT-A and TMT-B test scores.

Search strings in SH may be ascending or descending alphabetical sequences (e.g. 'A,B,C,…' or '…,C,B,A'), ascending or descending numeric sequences (e.g. '1,2,3,…' or '…,3,2,1'), ascending or descending alphanumeric sequences (e.g. '1,A,2,B,…' or '…,B,2,A,1'), and English language words selected out of a fixed lexicon (e.g. 'H,O,R,S,E').

A marker in SH appears as a circle containing a single letter or number. The centers of the markers are arranged on the board in a 4x8 grid with a spacing of 80 pixels. We refer to the position of all the markers on the board in this grid as the *layout* of the markers. Markers may have a diameter of 63 pixels or 77 pixels. At the normal viewing distance of about 25 cm, the markers subtend approximately 3 degrees of visual angle. This size assures 100% recognition for subjects with corrected vision to 20/20. In any round, all the markers have the same *width* (diameter). SH has two types of markers: (1) targets, and (2) distractors. *Targets* are those markers containing a character that appears in the search string and that the subject must select in the course of completing the connect-the-dots task for the round. *Distractors* are markers that contain characters not appearing in the search string.

A SH round has a variable number of targets, typically about 4 to 10, as well as additional distractors. The board for each SH round is generated at random. In order to track accurately the subject's performance over time, SH has a particular type of test pattern that appears regularly and often to serve as a baseline reference on a subject's performance over time; these rounds have the search string '1,2,3,4', and no distractors. These rounds make up about one in four SH rounds.

SH was designed to try to make the repetition of a very simple task as interesting to the subject as possible. The use of a smaller number of targets was believed to make the game faster, and the variability of the number of targets together with additional distractors was believed to add more variety to the game. Ascending numeric and alphanumeric sequences were included to facilitate comparison to TMT, and descending numeric and alphanumeric sequences as well as ascending and descending alphabetic sequences and lexical sequences were included to add further variety to the game.

*C. Differences Between SH & TMT*

While SH was designed to mimic TMT, the two tasks are clearly not identical, with differences including: (1) SH is played in-home at the subject's leisure, while TMT is an in-clinic test, (2) SH is a computer game while TMT is a pen-and-paper test, (3) a SH round has a 30 sec time limit while the subject is instructed to complete TMT-A or TMT-B as quickly as possible, (4) in SH the search string remains visible to the subject for the duration of the round while in TMT the subject is told the search string verbally before beginning the test, (5) the presence of the words "Begin" and "End" on TMT, (6) that SH marks a selected target by changing the color to green where TMT marks a selected target by trail passing through it, (7) SH boards can contain both targets and distractors while TMT contains only targets, and (8) SH board typically contain about 4-10 markers while TMT always contains 25. In Sec. III.A – III.D, we develop a model of the connect-the-dots task that is intended to produce a set of performance measures for each subject based on analysis of moves made playing SH. It is expected that these performance measures would be related to comparable measures based on analysis of moves made in TMT (were the move data available). However, comparison of the resulting performance measures between SH and TMT is complicated by the differences listed here. Instead of assuming that the performance measures are the same in the two cases, we suppose some set of transformations exist relating each performance between the cases of SH and TMT, and, for simplicity, that these transformations are the approximately the same for all subjects. These transformations are developed in detail in Sec. III.E, and given in Eq. (4).

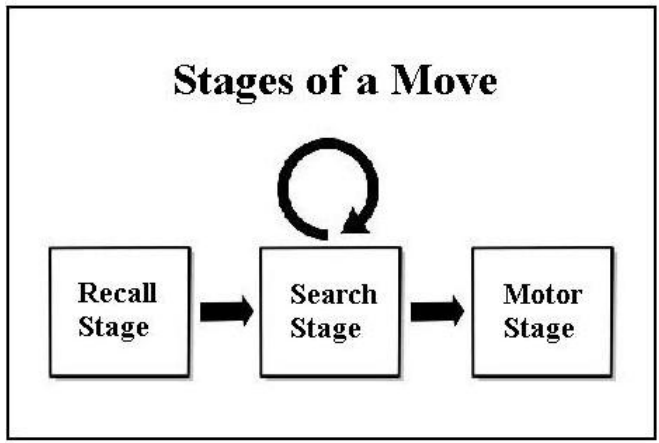

Fig. 3. Additive stages move model. The process of selecting the next target in the sequence involves three sequential stages of (1) recalling the next target, (2) serially searching for the next target by considering the available targets one after another, and (3) physically moving the mouse so that the cursor is on the target and clicking.

## III. Connect-the-Dots Model

The connect-the-dots task is completed by drawing a single line through a sequence of "dots" in a specified order. On a high level, performance on the connect-the-dots task can be characterized by the time taken to complete the entire task and the number of errors made in the course of completing the task (as is done in TMT); however, we choose to characterize subject performance by characterizing the performance across each individual move the subject makes in the course of completing the task.

The particular goal in the present paper is to show that measuring subject performance on moves observed in SH play can be used to estimate the subject's scores on TMT. The connect-the-dots model we develop in this section describes each move in the connect-the-dots task. By applying the model to the SH we can take all the observed moves from SH rounds and estimate a set of parameters characterizing how a subject makes a move in connect-the-dots tasks like those in SH. Conversely, given a set of parameters characterizing how a subject makes a move during TMT, we can construct an estimator of the TMT score in the case where no errors are made.

For the sake of simplicity, the proposed model does not characterize errors or the process of recovering from errors; consequently, in our analysis, we ignore the small proportion of rounds of SH in which the subject made any error. We ignore whole rounds to avoid any affects of the error on other moves, whether the error was due to something happening during an earlier move, or caused the subject to carry out subsequent moves differently than they otherwise would. Unfortunately, subjects do make errors on TMT, and we have to account for those errors in the estimators of the TMT-A and TMT-B scores (see Sec. VI for information on the numbers of errors made). We account for the observed errors by estimating the time delays due to the errors and including these in the prediction of the TMT-A and TMT-B scores. This approach is useful in estimating the relative contribution of the correctly executed moves and errors in the final test scores.

The connect-the-dots model decomposes a move into a sequence of three statistically independent stages as shown in Fig 3: (1) the *recall & update* stage during which the subject calls to mind the next target in the search string, (2) the *search* stage during which the subject searches among the unselected targets game board to locate the current target, and (3) the *motor* stage during which the subject moves the mouse or pen to the located target to select it. The general methodology corresponds to Donders' additive stages. [9, 10]

The statistical independence is based on the idea that each stage is affected by different aspects of the task and that the effect is limited to that stage. We expect that the duration of the recall & update stage would vary with the type of the search string (i.e. it should take a different amount of time to recall the next target when the search string is purely alphabetic or numeric as opposed to an alphanumeric search string). The duration of the search stage should depend on the number of additional distractors and unselected targets on the board, with the time spent in search decreasing on average as the subject moves to the end of the round. [21] Finally, the length of the motor stage should depend only on the distance on the board from the previously selected target to the new target – assuming that the target size is constant.

We now describe the detailed characterization of the stages of the model: recall & update, search and motor.

### A. Recall & Update Stage

The recall & update stage is characterized by the *recall time* $T_R$ required by the subject to recall the next target in the search string. The recall time is a random variable (RV) with expected value $\langle T_R \rangle = \tau_R$, and some standard deviation. We suppose that the time $T_R$ spent by the subject recalling the next target in a sequence may vary across the classes of search strings available in SH (i.e. alphabetic, numeric, alphanumeric, and lexical), but is assumed to be the same for all the targets in sequences of a given class. We denote values for $\tau_R$ intended to estimate recall for TMT-A-like connect-the-dots task by $\tau_R^A$, and for TMT-B-like tasks by $\tau_R^B$.

*B. Search Stage*

The search stage is characterized by the *search time* $T_S$ required by the subject to locate the next target in the search string after it has been recalled. We treat search as a series of discrete steps, [21] where the total number of steps in any search is a RV. In each step, the subject considers a different marker (a target or a distractor) on the board. The subject compares the marker being considered during that step to the target being searched for. If the subject decides that the marker being considered is the same as the target they are searching for, they select the marker; otherwise the subject continues the search to another step and considers another marker. Each step of search takes some fixed time $\tau_S$. We suppose that the time $\tau_S$ spent by the subject on each step of search may vary across the classes of search strings available in SH (i.e. alphabetic, numeric, alphanumeric, and lexical), but is assumed to be the same for all the targets in sequences of a given class. We denote values for $\tau_S$ intended to estimate search step time for TMT-A-like connect-the-dots task by $\tau_S^A$, and for TMT-B-like tasks by $\tau_S^B$.

The number of steps in a given search depends on the number of markers remaining on the board (i.e. the initial number of markers less the number of targets that have been selected thus far). We suppose that the subject searches the remaining markers at random, with no memory of any of the remaining markers from searches made in previous moves in the same round; we further suppose that during the search stage, the subject has perfect memory and considers each marker only once (we discuss the validity and utility of these assumptions in Sec. VII). Let us consider a SH round with $n$ targets and $d$ distractors. Suppose the subject is searching for the $v-th$ target. The subject has already found $v-1$ targets, so there are $n-v+1$ targets still on the board. The expected number of steps for the search is $(n-v+d+1)/2$. The expected value for the total search time $T_S$ for this target is given by:

$$\langle T_S \rangle = ((n-v+d+1)/2)\tau_S. \qquad (1)$$

The distribution of $T_S$ for a given value $n-v+d$ is uniform on the discrete values $\tau_S, \ldots, (n-v+d+1)\tau_S$.

*C. Motor Stage*

The motor stage is characterized by the *motor time* $T_M$, required by the subject move the mouse or pen from one target to the next. We suppose it to be independent of the search string. The movement made by the mouse or pen is a rapid movement into a target area given by the size of the marker being selected, and is expected to be consistent with Fitts' law. [11-13] We treat the motor time as a RV whose mean value satisfies Fitts' law and has some standard deviation. Fitts' law expresses the relationship between the distance $D$ from the initial position to the center of the target, the target width $W$, and the expected motor time, $T_M$, required to complete the move. Defining $D_v$ to be center-to-center distance between the $v-1-th$ and $v-th$ targets, assuming a common width $W$ for all targets, the expected motor time taken to move from the $v-1-th$ to the $v-th$ target is given by:

$$\langle T_M \rangle = a + b\log_2(D_v/W+1). \qquad (2)$$

The value $\log_2(D/W+1)$ provides a measurement of the amount of information the subject must process to complete the movement as measured in bits; so the value $b$ provides a measure of how much time the subject spends processing each bit of information.

*D. Total Time*

The expected time needed to complete an error-free connect-the-dots task with $n$ targets and $d$ distractors is simply the sum of the expected times for all of the component moves (we use the dot to distinguish multiplication $a \cdot (b)$ from the expression of a function $a(b)$):

$$\begin{aligned}
T &= n \cdot (\tau_R + a) + \kappa(n,d)\tau_S + \chi(D_1,\ldots,D_v,W)b, \\
\kappa(n,d) &= \sum_{v=1}^{n}((n-v+d+1)/2), \\
\chi(D_1,\ldots,D_v,W) &= \sum_{v=1}^{n}\log_2(D_v/W+1).
\end{aligned} \qquad (3)$$

We can see that the expected time required to complete a connect-the-dots task is linear in the parameters characterizing the subject's cognitive and motor abilities: $\tau_R$, $\tau_S$, $a$, and $b$. We call $\kappa(n,d)$ and $\chi(\{D_v\},W)$ the *search complexity*, and the *motor complexity* respectively.

Due to the way in which $\tau_R$ and $a$ appear in Eq. (3), their values cannot not be estimated separately. Instead, the best we can do is to estimate their sum $\tau_R + a$.

*E. Relating SH to TMT*

Due to the differences between SH and TMT outline in Sec. II.C, we do not expect the values $\tau_R$, $\tau_S$, $a$, and $b$ to relate trivially to their counterparts $\tau_R{'}$, $\tau_S{'}$, $a'$, and $b'$. Differences 7 and 8 from Sec. II.C, the presence of distractors and the numbers of targets, have already been handled in the model developed in this section. We suppose that the differences between SH and TMT do not affect the search or motor stages, so difference 6 regarding how selected targets are indicated is assumed not to affect search, and difference 2 regarding SH being a computer game and TMT being a pen-and-paper test is assumed not to effect movement from one target to the next. The remaining differences – 1, in-home versus in-clinic, 2, presence of time limit, 4, visibility of the search string, and 5, the presence of the words "Begin" and "End" – are assumed to only affect recall. We model the net effect of these differences on recall using the linear transformation $\tau_R' = \alpha + \beta \tau_R$. However, as a practical matter, we cannot separate the values $\tau_R$ and $a$, but rather we must work with $\tau_R + a$, so we use the approximate transformation $\tau_R' + a' \approx \alpha + \beta \cdot (\tau_R + a)$. The full set of transformations is:

$$\begin{aligned}\tau_R' + a' &\approx \alpha + \beta \cdot (\tau_R + a), \\ \tau_S' &= \tau_S, \\ b' &= b.\end{aligned} \quad (4)$$

This set of transformation is assumed to be the same for all search strings (i.e. the values $\alpha$ and $\beta$ are the same when relating SH to TMT for TMT-A-like search string and for TMT-B-like search strings). We also assume that our subject population is sufficiently homogeneous that we can use the same transformation for every subject.

## IV. ANALYZING SH PLAY

We validated the model of connect-the-dots tasks developed in Sec. III by constructing an estimator of the subject's scores on the TMT-A and TMT-B using measurements taken from that subject's play of SH. We used SH data from rounds using ascending and descending alphabetic and numeric search string to construct the TMT-A estimator and data from rounds using ascending and descending alphanumeric search string to construct the TMT-B estimator. We chose to combine round data in this way so that more data would be available for each subject and we would be able to retain as many subjects as possible for analysis (see Sec. VI for more information).

We now consider how to estimate a subject's cognitive and motor parameters $\tau_R^A + a$, $\tau_R^B + a$, $\tau_S^A$, $\tau_S^B$, and $b$ using the SH move data which consists only of timestamps indicating when buttons were selected by the subject and the relative positions of the buttons on the board. We produce the estimates in a two-step process. In the first step, we estimate the subject's motor performance as described by the Fitts' law $b$ parameter, from the time and position data, using SH rounds with search string '1,2,3,4' and no distractors using the model developed in Sec. III. In the second step, we use the estimated motor performance from the first step to remove the effect of motor performance from observed moves in SH rounds with alphabetic, numeric, and alphanumeric search strings (i.e. rounds with search strings of the same classes as that in the TMT-A and TMT-B respectively), and then estimate the subject's cognitive recall and search parameters $\tau_R^A + a$, $\tau_R^B + a$, $\tau_S^A$, and $\tau_S^B$ also using the model developed in Sec. III.

### A. Estimating Motor Parameters

The first step in our two step process of estimating a subject's cognitive and motor parameters is to estimate the subject's Fitts' law motor parameter $b$. We use a data set consisting only of moves from SH rounds with the search string '1,2,3,4' and no distractors. As there is some uncertainty in the position of the mouse at the beginning of the round, we ignore the move to the first target for each round.

For each move, we know the inter-target distances $D_i$, target widths $W_i$, observed move times $t_i$, numbers of targets $n_i$, and the position of the target in the search string $v_i$ (so for the string '1,2,3,4', the target '1' has $v = 1$, the target '2' has $v = 2$ and so on). So, for a particular move, the expected total time taken to move is:

$$\begin{aligned}\langle t_i \rangle &= (\tau_R + a) + ((n_i - v_i + 1)/2)\tau_S \\ &\quad + b \log_2(D_i / W_i + 1).\end{aligned} \quad (5)$$

As the values $\tau_R + a$, $\tau_S$, and $b$ are unknown at this point, we have to fit all three to the data. We can estimate their values by finding the values $c_0$, $c_1$, and $c_2$ that minimize the total squared error given by:

$$\begin{aligned}\varepsilon = \sum_i \big(&\langle t_i \rangle - c_0 - |c_1|((n_i - v_i + 1)/2) \\ &- |c_2| \log_2(D_i / W_i + 1)\big)^2.\end{aligned} \quad (6)$$

We constrain the result so that $c_1$ and $c_2$ are non-negative. From this step, we only retain the estimated value $b = |c_2|$.

### B. Estimating Cognitive Parameters

The first step in our two step process of estimating a subject's cognitive and motor parameters is to use the subject's Fitts' law motor parameter $b$ estimated in the first step to remove the motor component of the move time and estimate the subject's cognitive parameters $\tau_R^A + a$, $\tau_R^B + a$, $\tau_S^A$, and $\tau_S^B$. For estimation of $\tau_R^A + a$ and $\tau_S^A$, we use a data set consisting only of moves from SH rounds with the ascending or descending alphabetic or numeric search strings excluding SH rounds with search string '1,2,3,4' and no distractors; and for estimation of $\tau_R^B + a$ and $\tau_S^B$, we use a data set consisting only of moves from SH rounds with ascending or descending alphanumeric search strings. As there is some uncertainty in the position of the mouse at the beginning of the round, we ignore the move to the first target for each round.

As we did above in Eq. (5) we estimate the expected time taken for each move by summing the expected times for each of the three additive stages. In this case, however, we must also include values for the numbers of distractors $d_i$ present for each move, giving the expected move time:

$$\begin{aligned}\langle t_i \rangle &= (\tau_R + a) + ((n_i - v_i + d_i + 1)/2)\tau_S \\ &\quad + b \log_2(D_i / W_i + 1).\end{aligned} \quad (7)$$

As the values $\tau_R + a$ and $\tau_S$ are unknown at this point, we have to fit both to the data. We can estimate their values by finding the values $c_0$ and $c_1$ that minimize the total squared error given by:

$$\varepsilon = \sum_i \left( \langle t_i \rangle - c_0 - |c_1| \left( (n_i - v_i + d_i + 1)/2 \right) \right.$$
$$\left. - b \log_2 \left( D_i / W_i + 1 \right) \right)^2. \tag{8}$$

We constrain the result so that $c_1$ is non-negative. From this step, we retain the estimated values $\tau_R + a = c_0$ and $\tau_S = |c_1|$.

## V. ESTIMATING TMT SCORES

Using the procedure given in Sec. IV, we can produce estimates for the cognitive and motor parameters $\tau_R^A + a$, $\tau_R^B + a$, $\tau_S^A$, $\tau_S^B$, and $b$ for each subject using data from play of SH. We now use these estimates to produce estimators of performance on TMT-A and TMT-B for each subject.

### A. TMT Score Estimator

We begin our construction of the estimators for the performance on TMT by assuming we had the actual values of the cognitive and motor parameters $\tau_R^{A\prime} + a'$, $\tau_R^{B\prime} + a'$, $\tau_S^{A\prime}$, $\tau_S^{B\prime}$, and $b'$ that describe the subject's performance on TMT. Let us consider first the time spent completing TMT-A or TMT-B after the first target has been selected. The search complexity $\tilde{\kappa}$ for this portion of the test can be calculated from the definition in Eq. (3); while the motor complexities $\tilde{\chi}^A$, and $\tilde{\chi}^B$ for this portion of the test can be found using the definition in Eq. (3) and direct measurement of the layout of the markers on the test page. Using the superscript $X$ as a place-holder for either $A$ or $B$, indicating whether TMT-A or TMT-B is being considered, the expected total time spent completing the test after the first target has been found when no errors are made is:

$$\langle T^X \rangle = 24 \left( \tau_R^{X\prime} + a' \right) + \tilde{\kappa} \tau_S^{X\prime} + \tilde{\chi}^X b'. \tag{9}$$

One aspect of difference 2 in Sec. II.C is that the subject begins the test by turning over the test page. This adds some amount of amount $T^0$ to the final time. We suppose $T^0$ is the same for all subjects for both parts of TMT. In addition, the move to the first target begins at some unknown position. We suppose that the motor portion of the time taken to make the move to the first target is about average for the motor times on the test, or $a + (1/24) \tilde{\chi}^X b'$. When the first move is included, the search complexity is now $\kappa$ as given in Eq. (3) for the full test rather than $\tilde{\kappa}$. The expected total time spent completing the entire test when no errors are made is:

$$\langle T^X \rangle = T^0 + 25 \left( \tau_R^{X\prime} + a' \right) + \kappa \tau_S^{X\prime} + (25/24) \tilde{\chi}^X b'. \tag{10}$$

We denote the number of errors the subject made on the TMT-A and TMT-B respectively by $N^A$ and $N^B$ and treat time required by the subject to make and recover from an error as a random variable with mean $\theta$; we further assume that the random variable is the same for all subjects. Thus, the expected test score (or expected total time spent completing the test when errors are made) given a expected numbers of errors $\langle N^A \rangle$ and $\langle N^B \rangle$ is:

$$\langle S^X \rangle = \langle T^X \rangle = T^0 + 25 \left( \tau_R^{X\prime} + a' \right) + \kappa \tau_S^{X\prime}$$
$$+ (25/24) \tilde{\chi}^X b' + \theta \langle N^X \rangle. \tag{11}$$

Finally, we must replace the cognitive and motor parameters $\tau_R^{X\prime} + a'$, $\tau_S^{X\prime}$, and $b'$ describing performance on TMT by their counterparts that have been estimated from SH. Replacing the TMT cognitive and motor parameters $\tau_R^{X\prime} + a'$, $\tau_S^{X\prime}$, and $b'$ by their SH counterparts $\tau_R^X + a$, $\tau_S^X$, and $b$ using the transformation between the two connect-the-dots tasks is given in Eq. (4) gives the estimator for expected scores on TMT given expected numbers of errors:

$$\langle S^X \rangle = \left( T^0 + 25\alpha \right) + 25\beta \cdot \left( \tau_R^X + a \right) + \kappa \tau_S^X$$
$$+ (25/24) \tilde{\chi}^X b + \theta \langle N^X \rangle. \tag{12}$$

### B. Estimating Global Parameters

The subject-specific cognitive and motor parameters $\tau_R^A + a$, $\tau_R^B + a$, $\tau_S^A$, $\tau_S^B$, and $b$ appearing in Eq. (12) have been estimated using SH move data. The global parameters $T^0 + 25\alpha$, $\beta$, and $\theta$ (assumed to be the same for all subjects) now need to be estimated. We estimate the global parameters by finding the values of $T^0 + 25\alpha$, $\beta$, and $\theta$ that cause our estimators given in Eq. (12) to have optimal performance combined for both TMT-A and TMT-B across all subjects.

We index our subjects so that every subject has some index $i$ in the data related to TMT-A, and some index $j$ in that related to TMT-B. For each subject we have measurements average test scores $\langle S_i^A \rangle$ and $\langle S_j^B \rangle$, and the average numbers of errors made in each part $\langle N_i^A \rangle$ and $\langle N_j^B \rangle$. As the values $T^0 + 25\alpha$, $\beta$, and $\theta$ are unknown at this point, we have to fit all three to the data. We can estimate their values by finding the values $c_0$, $c_1$, and $c_2$ that minimize the total squared error given by:

$$\varepsilon = \sum_i \left( \langle S_i^A \rangle - c_0 - 25|c_1| \cdot \left( \tau_R^A + a \right)_i \right.$$
$$\left. - \kappa \tau_{S,i}^A - (25/24) \tilde{\chi}^A b_i - |c_2| \langle N_i^A \rangle \right)^2$$
$$+ \sum_j \left( \langle S_j^B \rangle - c_0 - 25|c_1| \cdot \left( \tau_R^B + a \right)_j \right. \quad (13)$$
$$\left. - \kappa \tau_{S,j}^B - (25/24) \tilde{\chi}^B b_j - |c_2| \langle N_j^B \rangle \right)^2.$$

We constrain the result so that $c_1$ and $c_2$ are non-negative. We retain the estimated values $T^0 + 25\alpha = c_0$, $\beta = |c_1|$, and $\theta = |c_2|$.

## VI. EMPIRICAL STUDY

30 older adults (25 female and 5 male, average age 80 ± 6.0 years, average level of education 15 ± 2.7 years, MMSE = 28 ± 1.1, ADL = 0.071 ± 0.30) participated in a one year study in which a set of computer games that included SH was placed into their homes.

SH was developed along with 8 other adaptive computer games to measure cognitive performance of individuals on a regular basis by monitoring their computer interactions during game play on their home computers. [22, 23] The set of computer games, including SH, was placed in subjects' homes for a period of one year. Subjects were encouraged to play all the games often, but were free to play the games as little as they wanted. Play of the computer games by the subjects was monitored, and the relevant information needed to reconstruct a subject's play in any of the games was recorded in a central database in a format allowing us to reconstruct any round of SH played. Subjects were given a battery of cognitive tests, including TMT, administered by trained clinical staff according to standard administrations procedures, at the beginning of the study, 6 months into the study and at the end of the study.

We restricted the analysis to those subjects for whom, when only error-free rounds of SH with alphabetic, numeric, and alphanumeric search strings were considered, we could find at least 25 moves total (across all such rounds) for rounds using alphabetic or numeric and at least 25 moves total for rounds using alphanumeric search strings. Data from SH rounds with alphabetic and numeric search strings were combined in the analysis. Rounds with search string '1,2,3,4' and no distractors, were excluded, as were the first moves within each round for the purpose of determining whether a subject had enough data. This restriction left a cohort of 23 older adults (20 female and 3 male, average age 81 ± 6.8 years, average level of education 15 ± 2.9 years, MMSE = 28 ± 0.89, ADL = 0.058 ± 0.16).

The numbers of moves across the remaining cohort 23 subjects for SH rounds with the search string '1,2,3,4' and no distractors ranged from 24 to 1236 (median of 108), for rounds with alphabetic or numeric search strings (not including moves from rounds with search string '1,2,3,4' and no distractors) the numbers of moves ranged from 35 to 4618 (median of 259), and for rounds with alphanumeric search string ranged from 43 to 4819 (median of 273). Data for rounds ascending and descending alphabetic or numeric search strings were pooled together for estimating TMT-A performance as were data for ascending and descending alphanumeric search strings for estimating TMT-B performance so that as many subjects as possible could be retained for analysis.

Following the first step of our two step procedure for estimating subject cognitive and motor parameters from SH data given in Sec. IV, for each subject we estimated the value for the Fitts' law parameter $b$ (Eq. (2)) by minimizing the total error expressed in the model given in Eq. (6) using all observed error-free SH rounds with search string '1,2,3,4' and no distractors. The observed means and standard deviations for the values of $R^2$ and p for the fit of the model given in Eq. (6) across the 23 subjects were $R^2 = 0.26 ± 0.11$ and p = 0.0080 ± 0.028, and the mean and standard deviation of the numbers of moves available for each subject for the estimation was n = 590 ± 810. The observed mean and standard deviation of the estimated Fitts' law parameter $b$ across the full cohort of 23 subjects was:

$$b = 300 \pm 110 \, \text{ms/bit}. \quad (14)$$

The SH rounds chosen for the estimation of the Fitts' law parameter $b$ were intended to be those for which the motor component would be strongest. The model fit all subjects at a significance level of $p < 0.05$. The low $R^2$ values are expected due to the uniform distribution of the number of search steps given a number of unselected targets on the board described in Sec. III.B. Our average value for $b$ is close to the independently measured value of 166 ms/bit measured for point-select methods of selecting icons in a computer interface, [24] with the measured value being about one deviation below our average $b$. We consider this further in Sec. VII.

Continuing to the second step of our two step procedure for estimating subject cognitive and motor parameters from SH data, we estimated the recall and search performances $\tau_R^A + a$, $\tau_R^B + a$, $\tau_S^A$, and $\tau_S^B$ for each subject using the previously estimated values of $b$ by minimizing the total error expressed in the model given in Eq. (8) using data from alphabetic and numeric search strings to estimate $\tau_R^A + a$ and $\tau_S^A$, and alphanumeric search strings to estimate $\tau_R^B + a$ and $\tau_S^B$. The observed means and standard deviations for the values of $R^2$ and p for the fit of the model given in Eq. (8) across the 23 subjects for estimation of the parameters $\tau_R^A + a$ and $\tau_S^A$ were $R^2 = 0.097 ± 0.061$ and p = 0.065 ± 0.17 with the number of moves available use in the estimation having mean and standard deviation n = 670 ± 1200, and for estimation of the parameters $\tau_R^B + a$ and $\tau_S^B$ were $R^2 = 0.087 ± 0.050$ and p = 0.049 ± 0.12 with the numbers of moves available use in the estimation being n = 730 ± 1300. In the case of the estimation of $\tau_R^A + a$ and $\tau_S^A$, four of the subjects had fits with p > 0.05, and among these subjects the numbers of moves available for estimation had means and standard deviations of n = 52 ± 19; similarly for the case of the estimation of $\tau_R^B + a$ and $\tau_S^B$, four of the subjects had fits with p > 0.05, and among these subjects the numbers of moves available for estimation had means and standard deviations of n = 82 ± 38. Two subjects had fits with p > 0.05 in both cases. Removing the appropriate four subjects with poor fits in the each of the two cases gives, for the remaining 19 subjects, in the former case $R^2$ = 0.11 ± 0.057, p = 0.0032 ± 0.0077, and n = 800 ± 1300, and, for the remaining 19 subjects, in the latter case $R^2$ = 0.10 ± 0.043, p = 0.0039 ± 0.0095, and n = 830 ± 1300. The observed means and standard deviations of the estimated cognitive parameters across the full cohort of 23 subjects were:

$$\tau_R^A + a = 380 \pm 250 \, \text{ms},$$
$$\tau_R^B + a = 660 \pm 280 \, \text{ms},$$
$$\tau_S^A = 96 \pm 43 \, \text{ms},$$
$$\tau_S^B = 110 \pm 41 \, \text{ms}.$$
(15)

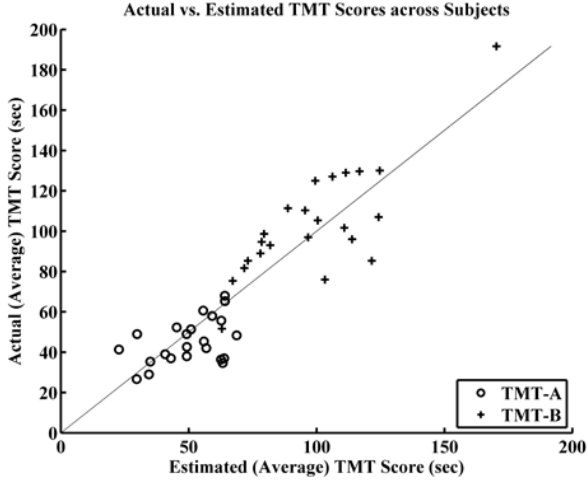

Fig. 4. Actual vs. Estimated TMT Scores across Subjects. Each of the 23 subjects has two values shown, one for TMT-A and one for TMT-B, each representing the average of the three administrations of TMT. The model fit has $R^2 = 0.82$ and $p < 0.0001$. A line with slope one passing through the origin is shown for reference.

The model, again, fit most subjects to a statistical significance level of p < 0.05, and the cases where this level of significance was not met appear to be attributable to the smaller amount of data available. Again, the low $R^2$ values are expected due to the uniform distribution of the number of search steps given a number of unselected targets and distractors on the board described in Sec. III.B. Due to the fact that we are measuring combined cognitive and motor values $\tau_R^A + a$ and $\tau_R^B + a$ rather than the purely cognitive values $\tau_R^A$ and $\tau_R^B$, we could not compare the measured values to existing research. However, we could estimate the set-switching (the time needed for the subject to switch from considering the sequence of numbers to considering the sequence of letters and vice versa in TMT-B) by taking the difference of $\tau_R^A + a$ and $\tau_R^B + a$. The average estimated set-switching time of 280 ms compared well with independently measured values of about 200 ms. [25] The average estimated values for $\tau_S^A$ and $\tau_S^B$ for each step in visual search differed from the independently measured value of 240 ms [26] by about a factor of two. We consider these further in Sec. VII.

The observed TMT-A and TMT-B scores and numbers of errors across all the tests taken by the subjects being included in this analysis and their standard deviations were $S^A = 45 \pm 11$ s and $N^A = 0.0073 \pm 0.14$, and $S^B = 100 \pm 28$ s and $N^B = 1.0 \pm 0.64$. The observed scores are near those given in Sec. II.A (i.e. [18]) for subjects around the age and education of those used in our study. The observed TMT test-retest reliability for the test pairs: (1) beginning and 6 months, (2) beginning and 1 year, and (3) 6 months and 1 year, was observed to have $R^2$ of 0.32, 0.20, and 0.13 for TMT-A, and $R^2$ of 0.43, 0.30, and 0.59 for TMT-B. As we used the full year's worth of data to estimate subject performance, we characterized subject performance on TMT using averages of the test scores over the year.

We next constructed estimators of the mean TMT scores given the mean numbers of errors made on the tests using the procedure given in Sec. V. The motor complexities for the TMT-A and TMT-B for all moves after the first target has been selected were measured from the test pages using a ruler, giving $\tilde{\chi}^A = 66$ bits, and $\tilde{\chi}^B = 74$ bits. Using the values of the cognitive and motor parameters $\tau_R^A + a$, $\tau_R^B + a$, $\tau_S^A$, $\tau_S^B$, and $b$ that we estimated for the 23 subjects (Eqs. (14) and (15)), we estimated the global parameters $T^0 + 25\alpha$, $\beta$, and $\theta$ by minimizing the total error expressed in the model given in Eq. (13) using, for each subject the means of the three test scores and the means of the numbers of errors made in the course of each test administration. The model was fit with $R^2 = 0.82$ and $p < 0.0001$, and the estimated global parameter values were:

$$T^0 + 25\alpha = -9.1 \, \text{s},$$
$$\beta = 2.2, \quad (16)$$
$$\theta = 30 \, \text{s}.$$

For comparison, we looked at the performance of a simple linear regression of the test scores on the number of errors; the fit in this case had $R^2 = 0.58$ and $p < 0.0001$.

Inspection of the 95% confidence intervals for the coefficient estimates showed that the estimates of $\beta$ and $\theta$ were statistically significant, while that of $T^0 + 25\alpha$ was not. The values in Eq. (16) suggested that subjects required 30 sec to recover from an error. We consider this further in Sec. VII. In Fig. 4, we show how the estimated average TMT-A and TMT-B test scores using these values of the global parameters compared to the actual average test scores for each subject.

It was of interest to see how the model developed in this paper would perform in the case where no errors were made on TMT. We restricted the analysis to only include administrations of TMT in which both TMT-A and TMT-B had no errors. There were 16 subjects who had at least one error-free administration of TMT. We fitted a truncated version of our model in Eq. (13) lacking the terms in $\theta N^A$ and $\theta N^B$, and used the average of all error-free TMT administrations for the test scores. The model fit with $R^2 = 0.55$ and $p < 0.0001$, and estimated global parameter values of $T^0 + 25\alpha$ = -9.1 s and $\beta$ = 3.1. Inspection of the confidence intervals for the coefficient estimates showed that the estimate of $\beta$ was statistically significant while that of $T^0 + 25\alpha$ was not. The results were close to those found by retaining tests where errors were made in Eq. (16).

If the subject whose data give the outlier is removed from the data set, and the procedure repeated, the fit became $R^2$ = 0.73 and p < 0.0001, and the global parameters were found to be $T^0 + 25\alpha$ = -6.3 s, $\beta$ = 2.0, and $\theta$ = 29 s, with $\beta$ and $\theta$ being statistically significant while $T^0 + 25\alpha$ was not. These results were very close to those found when retaining the outlier in Eq. (16), so we retained the outlier in our analysis. The linear regression of the test scores on the number of errors had $R^2$ = 0.51 and p < 0.0001. When we limited ourselves to error-free administrations of the TMT, we were left with the same set of 16 subjects as in the previous paragraph.

## VII. Discussion

The connect-the-dots model developed in this study is a very simple model of the connect-the-dots task, and incorporates a number of simplifying assumptions, particularly in the search stage. Our simple model relating the measurements from SH to the case of TMT using a single set of transformations taken to be approximately the same for all subjects is based on the assumption of a relatively homogeneous set of subjects. A broader range of subjects may need to be grouped into classes each with a different set of transformations. A further limitation to our analysis presented here is the exclusion of SH rounds with errors, possibly limiting our ability to collect data for (mildly) cognitively impaired subjects.

We treated the time spent recalling the next target in the sequence as a single RV that appears once in each move. Because our empirical measurements did not allow to separate the effects of cognitive and motor segments, we were unable to estimate the recall times $\tau_R$, but had to measure a combined cognitive and motor parameter $\tau_R + a$. As a result, we could only average estimated set-switching time of 280 ms measured by taking the difference of $\tau_R^A + a$ and $\tau_R^B + a$ to independent measurements; though they compared well to the independently measured value of about 200 ms. [25] The model we developed to describe the transformation from the SH case to the TMT case should have taken the form $\tau_R' = \alpha + \beta\tau_R$, but, since we could only use values $\tau_R + a$, we had to make the approximation $\tau_R' + a' \approx \alpha + \beta \cdot (\tau_R + a)$ which causes the motor parameter $a$ to change value between the two cases. Even were this not a problem, the transformation $\tau_R' = \alpha + \beta\tau_R$ causes the set-switching time to change between the SH case and the TMT case, where it seems reasonable that the set-switching time would not change.

The model we use for visual search is that of a serial search that does not benefit from the memory from previous search stages for previous "dots," during any given round, but has perfect memory within the search stage for the current "dot."

Using this model, we estimated times spent on each step of search of about 100 ms in both cases ($\tau_S^A$ and $\tau_S^B$). The natural sequence of images produced by the eye during visual search has been independently measured to be about 240 ms per image. [26] This suggests that our model over-estimates the average number of steps in the visual search for a given target by about a factor of two. A more sophisticated model of search would include aspects related to: (1) visual acuity and how much target information the subject can take in at each search step, (2) the ability of the subject to remember target locations from previous searches, (3) the degree to which the subject becomes confused and considers the same target multiple times during a single search, and (4) the ability of the subject to ignore already selected targets and whether they spend much time considering these targets in later searches. What our simple model of search does capture is the observation that the total search times for moves are, on average, longer earlier in the connect-the-dots task than they are later in the task. [8]

We used Fitts' law to describe the motor portion of a move. An average value for the Fitts' law parameter of $b \approx 300$ ms/bit was measured across the subjects analyzed. We can compare this value to independently measured values for Fitts' law for several methods of using a computer mouse to select a icon: (1) $a$ = 135 ms and $b$ = 249 ms/bit for drag-select, (2) $a$ = 230 ms and $b$ = 166 ms/bit for point-select, and (3) $a$ = 135 ms and $b$ = 249 ms/bit for stroke-through. [24] Point-select should be closest to the button selection process happening in SH, so our average value of $b$ is roughly twice as large of the independently measured value (although that value is approximately one standard deviation below our average $b$).

We included the time spent recovering from an error as a single global parameter with the same value for all subjects. The estimated 30 s recovery time for each error is somewhat long (we do not have data on the duration of errors during the administrations of TMT). However, the average number of errors on TMT-A was near zero and that on TMT-B near one, so it is likely that the value estimated for $\theta$ largely reflects the time needed to recover from errors during TMT-B. The value for $\theta$ may well be inflated by a correlation where subjects making more errors also require more time to recover from each one. Alternatively, it is possible that, when taking the TMT, subjects also made a number of "near errors" in which they came close to selecting and incorrect "dot," but corrected the error themselves. If the number of these "near errors" is correlated to the number of actual errors, then the large error recovery time may reflect time taken up in a "near error" process as well.

Possibly related to variability in the numbers of errors between administrations of TMT and variability in the error-recovery time is the low test-retest reliability observed in Sec. VI. It appears from our results that errors can contribute a large amount of time to the test scores, and errors are discrete events, so differing numbers of errors from one test to another appear to be able to result in substantially different scores from one test to another.

Moving forward, it is important to better understand the process of making errors in the course of the connect-the-dots task, and there are a variety of approaches to doing this. There are four approaches one might take in considering the errors. (1) Look at the rate at which errors are made in playing SH and see whether this rate predicts the average number of errors made on TMT, or allows us to produce good estimates of the average TMT score without using the observed number of errors on TMT. (2) Look at how observed errors in SH relate to moves made immediately before and after. In our analysis, we have dropped all SH rounds in which any errors were made. The amount of data available, particularly for (mildly) cognitively impaired subjects, would be increased by dropping moves expected to be affected by observed errors from the data set rather than whole rounds. (3) Look at the moves in SH in which errors are made and see if the time spent recovering from the error can be used to predict error recovery times in TMT. (4) Look at outliers in the move times and see if the frequency and average duration of these outliers could be related to the number of errors or error recovery time in SH or TMT, suggesting that the outliers may be "near errors."

## VIII. Conclusion

The key objective of this work included (1) the development of techniques that would allow frequent assessment of cognitive functions of individuals at risk, for example associated with aging and (2) the demonstration of the utility of computational modeling in assessment of cognitive function. The general approach was based on using computer games that would enable neurophysiological assessments comparable to those obtained with traditional neuropsychological tests such as the TMT. In this study, we used a simple game that is similar to the TMT and was developed for this purpose as a part of prior study. [20, 22, 23] In addition to the estimation of the TMT performance, the objective of using the game was to derive a more refined assessment of the various cognitive components recruited in the execution of TMT.

The potential benefit of our approach to modeling computer interactions is that we can model cognitive performance over time for individuals in the home in a more scalable and less expensive manner than current standard practice. Cognitive measures and trends in these measures can be used to identify individuals for further assessment, to provide a mechanism for improving the early detection of neurological problems, and to provide feedback and monitoring for cognitive interventions in the home.